\documentclass[
	aps,pra,twocolumn, notitlepage,
	10pt,superscriptaddress
]{revtex4-1}
\pdfoutput=1
\usepackage{graphicx}
\usepackage{times,bbm,amsmath,amssymb}
\usepackage{xcolor}
\usepackage{hyperref}
\usepackage{color}
\usepackage{graphicx}
\usepackage{dcolumn}
\usepackage{bm}
\usepackage{soul}

\begin{document}

\title{Experimental adaptive Bayesian estimation of multiple phases with limited data}

\author{Mauro Valeri}
\affiliation{Dipartimento di Fisica, Sapienza Universit\`{a} di Roma, Piazzale Aldo Moro 5, I-00185 Roma, Italy}
\author{Emanuele Polino}
\affiliation{Dipartimento di Fisica, Sapienza Universit\`{a} di Roma, Piazzale Aldo Moro 5, I-00185 Roma, Italy}
\author{Davide Poderini}
\affiliation{Dipartimento di Fisica, Sapienza Universit\`{a} di Roma, Piazzale Aldo Moro 5, I-00185 Roma, Italy}
\author{Ilaria Gianani}
\affiliation{Dipartimento di Fisica, Sapienza Universit\`{a} di Roma, Piazzale Aldo Moro 5, I-00185 Roma, Italy}
\affiliation{Dipartimento di Scienze, Universit\`a degli Studi Roma Tre, Via della Vasca Navale 84, 00146, Rome, Italy}
\author{Giacomo Corrielli}
\affiliation{Istituto di Fotonica e Nanotecnologie, Consiglio Nazionale delle Ricerche (IFN-CNR), Piazza Leonardo da Vinci, 32, I-20133 Milano, Italy}
\affiliation{Dipartimento di Fisica, Politecnico di Milano, Piazza Leonardo da Vinci, 32, I-20133 Milano, Italy}
\author{Andrea Crespi}
\affiliation{Istituto di Fotonica e Nanotecnologie, Consiglio Nazionale delle Ricerche (IFN-CNR), Piazza Leonardo da Vinci, 32, I-20133 Milano, Italy}
\affiliation{Dipartimento di Fisica, Politecnico di Milano, Piazza Leonardo da Vinci, 32, I-20133 Milano, Italy}
\author{Roberto Osellame}
\affiliation{Istituto di Fotonica e Nanotecnologie, Consiglio Nazionale delle Ricerche (IFN-CNR), Piazza Leonardo da Vinci, 32, I-20133 Milano, Italy}
\affiliation{Dipartimento di Fisica, Politecnico di Milano, Piazza Leonardo da Vinci, 32, I-20133 Milano, Italy}
\author{Nicol\`o Spagnolo}
\affiliation{Dipartimento di Fisica, Sapienza Universit\`{a} di Roma, Piazzale Aldo Moro 5, I-00185 Roma, Italy}
\author{Fabio Sciarrino}
\email{fabio.sciarrino@uniroma1.it}
\affiliation{Dipartimento di Fisica, Sapienza Universit\`{a} di Roma, Piazzale Aldo Moro 5, I-00185 Roma, Italy}

\begin{abstract}
Achieving ultimate bounds in estimation processes is the main objective of quantum metrology. In this context, several problems require measurement of multiple parameters by employing only a limited amount of resources. To this end, adaptive protocols, exploiting additional control parameters, provide a tool to optimize the performance of a quantum sensor to work in such limited data regime. Finding the optimal strategies to tune the control parameters during the estimation process is a non-trivial problem, and machine learning techniques are a natural solution to address such task. Here, we investigate and implement experimentally for the first time an adaptive Bayesian multiparameter estimation technique tailored to reach optimal performances with very limited data. We employ a compact and flexible integrated photonic circuit, fabricated by femtosecond laser writing, which allows to implement different strategies with high degree of control. The obtained results show that adaptive strategies can become a viable approach for realistic sensors working with a limited amount of resources.
\end{abstract}

\maketitle

Quantum sensing devices are among the most promising quantum technologies. Their implementation relies on the use of quantum probes to attain enhanced performances in the estimation of one or more parameters compared to classical ones. Quantum metrology aims at identifying the best strategy able to provide this quantum advantage \cite{giovannetti2004quantum,giovannetti2006quantum,paris2009quantum, schnabel2010quantum, giovannetti2011advances, pezze2018quantum,pirandola2018advances}. This is achieved by carefully tailoring the probe state, the interaction, and the measurement, in order to extract the information on the relevant parameter, and by the optimal choice of the estimator through data post-processing \cite{Gianani19}. When performing a single parameter estimation, the optimal strategy is unequivocally identified through the saturation of the Cram\'er-Rao bound (CRB), which establishes the maximum achievable precision on the measured parameter \cite{helstrom1976quantum}. The CRB is asymptotically saturated with the number of resources employed to probe the system during the measurement. 
Conversely, the realisation of quantum sensors, able to perform estimations in realistic scenarios, poses two constraints to sensing devices: the resources to be used for probing are limited, and systems can show high complexity, often involving more than one parameter. 
Due to the finite number of available resources, it becomes of paramount importance to optimize the estimation protocols in order to reach the sought accuracy bounds employing the smallest number of resources possible. Such problem has been explored recently with several theoretical analysis. A possible approach towards the protocol optimization is that of exploiting adaptive strategies. These have been successfully employed in single-parameter estimation \cite{berry2000optimal,armen2002adaptive, wheatley2010adaptive,higgins2007entanglement,berni2015ab, paesani2017estimation, rubio2019limited, piccoloLume, daryanoosh2018experimental}. In this regard, machine learning (ML) approaches have provided a significant speed up in the saturation of the ultimate bounds \cite{hentschel2009adaptive, lovett2013differential, piccoloLume,palittapongarnpim2017learning}.

On the other hand, measuring multiple parameters at once might be necessary in complex systems characterized by a set of parameters, where a time or spatial dependency can prevent the successful realization of subsequent single-parameter estimations. 
The parameters considered can span from multiple phases \cite{polino2019experimental, peter13, mario17}, to phase and phase diffusion in frequency-resolved phase measurements \cite{genoni2012,mihai14, altorio15}, and phase and loss in absorbing systems \cite{Albarelli19}. 
In other instances where a system depends solely on one parameter, a multiparameter approach could still be favourable as other parameters can be interrogated as a control to monitor the quality of the sensor itself \cite{Roccia18,cimini1, cimini2}. In general, the saturable bounds for quantum multiparameter strategies are not as defined as in the single parameter case, and trade-offs in the achievable precision for each of the parameters have to be sought \cite{albarelli2019perspective, ragy2016compatibility, szczykulska2016multi,nichols2018multiparameter,gessner2019metrologicalmulti}.

In this context, it becomes of paramount importance to identify both a suitable estimation scenario and a corresponding  platform for an experimental investigation of adaptive multiparameter estimation protocols. A notable scenario to investigate is multiphase estimation  \cite{macchiavello2003optimal,humphreys2013quantum,ballester2004entanglement,liu2016quantum,gagatsos2016gaussian, mario17,ge2018distributed,ciampini2016scirep,gessner2018sensitivity,gatto2019distributed,guo2019distributed,polino2019experimental}. Not only such scenario provides a benchmark for multiparameter quantum metrology, but it has a plethora of practical applications in quantum imaging \cite{szczykulska2016multi,albarelli2019perspective}. A fundamental step is to find a suitable experimental platform to realize multiphase estimation. A viable solution is provided by integrated photonics, which enables the implementation of complex circuits with reconfiguration capabilities \cite{carolan2015universal, orieux2016recent, wang2018multi, atzeni2018source, taballione20198,wang2019integrated} with applications ranging from quantum simulation, to computation, and communication. This platform represents a promising system for optical quantum metrology, since interferometers with several embedded phases can be employed as a benchmark platform to study multiparameter estimation problems. In this direction, first results on multiphase estimation with quantum input probes have been recently reported \cite{polino2019experimental}, using a three-arm interferometer fabricated by femtosecond laser writing \cite{della2008micromachining, gattass2008femtosecond}.
Here, we report on multiphase estimation experiments performed with an integrated platform using different adaptive protocols. We identify the strategy providing better performances in terms of optimal estimation and computational resources. In particular, we have experimentally tested the proposed approach by feeding with single-photon states an integrated three-mode interferometer realized by the femtosecond laser writing technique.
The employed technique is a Bayesian learning protocol which exploits advantages of Montecarlo approach, such as its independence of integration space dimensions \cite{granade2012online}. This solution seems to be ideal for adaptive multiparameter problems, where complex optimizations could involve multiple integrations. 
Here we employ experimentally for the first time an adaptive strategy optimized in the limited data regime for multiphase estimation.  Using this algorithm, we demonstrate that the convergence to the CRB can be experimentally achieved after only few photons. Importantly, such convergence is achieved for both the simultaneously estimated phases. Our results improve research for identifying optimal learning strategy and finding experimental platforms suitable to test multiparameter estimation problems also in the limited data regime. 

\section*{Results}

\subsection*{Bayesian multiparameter estimation}

In multiparameter estimation, the aim is to measure simultaneously an unknown set of parameters $\bm{x} = (x_1, \ldots, x_n)$ by reaching the maximum precision allowed by the amount of resources employed in the process. In general, the set of parameters is encoded within the evolution of a system, either described through a unitary operator $U_{\bm{x}}$ or a more general map $\mathcal{L}_{\bm{x}}$. The value of the unknown parameters $\bm{x}$ can be estimated by preparing a suitable probe state $\rho$ and sending it to evolve throughout the system. Information on the unknown parameters can be retrieved by measuring the output state $\rho_{\bm{x}}$ with a set of measurement operators $\{\Pi_{d}\}$, where $d=1,\ldots,m$ represents the set of possible outcomes. Such process is then repeated $N$ times to improve precision in the estimation process. After $N$ probes have been prepared and measured, the obtained sequence of measurement outcomes $\bm{d} = (d_{1}, \ldots, d_N)$ has to be converted in a set of parameters estimates $\bm{\hat{x}}$ through a suitably chosen function $\bm{\hat{x}} = \bm{\hat{x}}(d_{1}, \ldots, d_N)$. A possible choice of estimator is provided by Bayesian protocols. This class of estimators is based on encoding the initial knowledge on the parameters in a probability function $p(\bm{x})$, called prior distribution, which is updated according to the Bayes rule at each step of the estimation protocol. The posterior distribution after $N$  probes reads $p(\bm{x} \vert \bm{d}) = \mathcal{N}^{-1} p(\bm{d} \vert \bm{x}) p(\bm{x})$, where $p(\bm{d} \vert \bm{x})$ is the likelihood function of the system expressing the conditional probability of obtaining the measurement sequence $\bm{d}$ for given values of the parameters $\bm{x}$, and $\mathcal{N}$ is a normalization constant. Then, the mean of the posterior distribution can be exploited as the estimate of the unknown parameters $\hat{x}_i = \int x_i p(\bm{x} \vert \bm{d}) \prod_{i} d x_i$. Bayesian protocols present several important properties. In particular, it can be shown that such approach is asymptotically unbiased, meaning that the estimated values converge to the true values when $N$ is large enough. This is related to the quadratic loss $L(\bm{x}, \bm{\hat{x}}; \bm{\tilde{w}}) = \sum_{i} \tilde{w}_{i} (x_{i} - \hat{x}_{i})^{2}$, whose average value over all measurement sequences $\bm{d}$ is commonly employed as a figure of merit to quantify the convergence of the estimation process. The coefficients $\tilde{w}_{i}$ can be chosen to reflect different weights between the parameters, while for equally relevant parameters they can be set as $\tilde{w}_{i} = 1$. Hereafter, we will consider this latter scenario and thus define the quadratic loss as $L(\bm{x}, \bm{\hat{x}}) = \sum_{i} (x_{i} - \hat{x}_{i})^{2}$. Furthermore, in a Bayesian framework the posterior distribution also provides a confidence region for the parameters estimates, which is represented by the covariance matrix $\mathrm{Cov}(\bm{\hat{x}})$ of $p(\bm{x} \vert \bm{d})$. This latter figure of merit is obtained for each single estimation experiment composed of a sequence of $N$ probes, and has no counterpart in frequentist approaches \cite{li2019bayes}. In general, Bayesian bounds for both the quadratic loss and the covariance matrix depend on the amount of a priori knowledge $p(\bm{x})$ available \cite{li2019bayes, rubio2018asympt, rubio2019limited,rubio2019bayesian}. Asymptotically for large values of $N$, corresponding to the regime where the amount of information acquired in the estimation process far exceeds the a priori knowledge, the covariance matrix satisfies the Cram\'er-Rao inequality $\mathrm{Cov}(\bm{x}) \geq \mathcal{F}^{-1}/N$, where $\mathcal{F}$ is the Fisher information matrix \cite{liu2019quantum} and thus $\mathcal{F}^{-1}$ corresponds to its inverse. Such quantity also provides an asymptotic bound for the quadratic loss as $L(\bm{x}, \bm{\hat{x}}) \geq \mathrm{Tr}[\mathcal{F}^{-1}]/N$.

\begin{figure}[ht!]
\centering
\includegraphics[width=.49\textwidth]{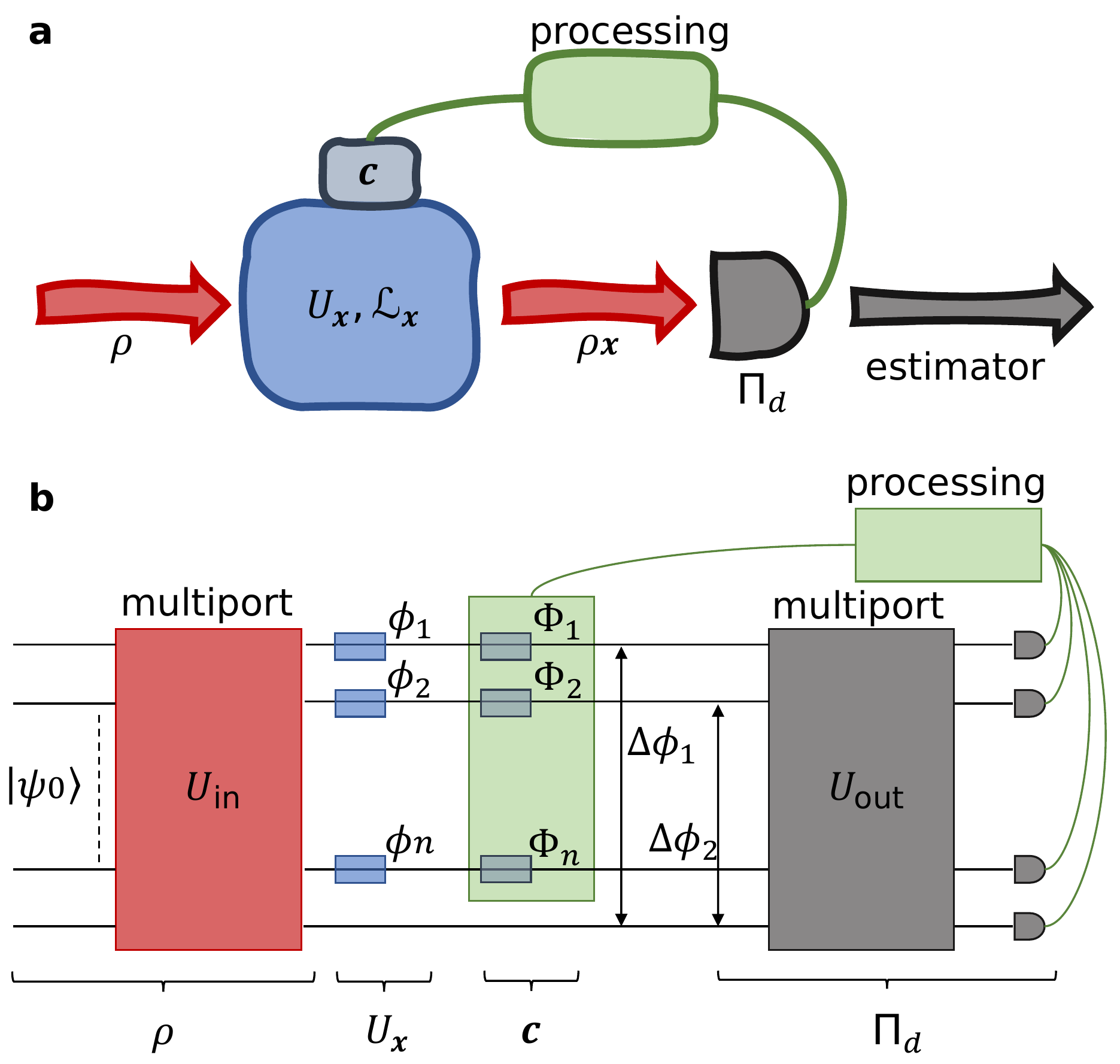}
\caption{{\bf a}, Scheme of an adaptive parameter estimation protocol. A sequence of probes $\rho$ are sent to estimate the parameters $\bm{x}$. At each step the results of the measurement $\Pi_d$ and the current knowledge on $\bm{x}$ are employed to optimize the control parameters $\bm{c}$. {\bf b}, Multiarm interferometer for multiphase estimation. An $m$-mode interferometer embeds $d = m-1$ unknown phase shifts $\bm{\phi}$, while additional controlled phases $\bm{\Phi}$ can be employed for adaptive protocols.}
\label{fig:concept} 
\end{figure}

Adaptive protocols can be employed when, besides the set of unknown parameters $\bm{x}$, the user has access to an additional set of control parameters $\bm{c} = (c_1, \ldots, c_l)$ that can be changed throughout the estimation process. More specifically, after each of the $N$ probes is sent and measured, the acquired knowledge is employed to change the values of $\bm{c}$ for the next probe to maximize the extraction of information in the subsequent measurement. Within a Bayesian framework, such knowledge is encoded in the posterior distribution. Hence, after each step of the estimation protocol, the user can decide the values of the control parameters $\bm{c}$ starting from $p(\bm{x} \vert \{ \bm{c} \}, \bm{d})$ (see Fig. \ref{fig:concept}a). Adaptive protocols represent a relevant tool in phase estimation process. Indeed, the adoption of adaptive strategies becomes a crucial requirement even in the single-parameter case to optimize the algorithm performances \cite{wiseman1995adaptive,berry2000optimal,armen2002adaptive,hentschel2009adaptive, wheatley2010adaptive, granade2012online, lovett2013differential, wiebe2016efficient, paesani2017estimation, piccoloLume}, with the aim of achieving the ultimate bounds provided by the Cram\'er-Rao inequality for small values of $N$ \cite{piccoloLume}. Furthermore, in more complex systems characterized by a phase-dependent Fisher information matrix, adaptive strategies become crucial to reach equal performances for all values of the unknown parameter(s) \cite{cimini2}.

\subsection*{Adaptive protocols for multiarm interferometers}

Given the general scenario described in the previous section, it is crucial to identify and test experimentally protocols to saturate the ultimate bounds with a very limited number of probes. In this context, multiarm interferometers represent a benchmark platform to perform simultaneous estimation of multiple phases. The platform is schematically shown in Fig. \ref{fig:concept}b, and represents the $m$-mode generalization of a Mach-Zehnder interferometer in the multimode regime \cite{spagnolo2012scirep, chaboyer2015scirep, ciampini2016scirep, polino2019experimental}. More specifically, it is composed by a sequence of a first multiport splitter, employed to prepare the probe state, a series of phase shifts between all the optical modes, and a second multiport splitter which defines the output measurement. Both multiport splitters can be in principle designed according to appropriate decompositions \cite{reck1994experimental, clements2016optimal} to implement any linear unitary transformation. The internal phase shifts can be divided in two layers. The first one $\bm{\phi} = (\phi_1, \ldots, \phi_n)$ corresponds to the unknown parameters to be measured, while the second one $\bm{\Phi} = (\Phi_1, \ldots, \Phi_n)$ takes the role of the control parameters for adaptive estimation; we note that in our implementation the number of controls $l=n$. Here, $n = m-1$ is the number of independent parameters, since one of the phases is considered as the reference mode. Both the unknown parameters and the control ones contribute to the overall phase differences $\bm{\Delta \phi} = (\Delta \phi_1, \ldots, \Delta \phi_n)$ within the interferometer.

We study different adaptive protocols for Bayesian learning of the unknown phases of this platform injected by a single-photon state, by focusing both theoretically and experimentally on the three-mode scenario ($m=3$) with two independent parameters ($n=2$). More specifically, we choose both for state preparation and state measurement transformation a balanced tritter described by unitary matrix $U$ with $\vert U_{i,j} \vert^2 = 1/3, \; \forall (i,j)$ \cite{spagnolo2013tritter}. Injecting a single photon on input port $1$ corresponds to generating a sequence of probe states of the form $\vert \psi_\mathrm{in} \rangle = 3^{-1/2} (\vert 1,0,0 \rangle + \vert 0,1,0 \rangle + \vert 0,0,1 \rangle)$, which represents a single-photon state exiting in the balanced superposition of the three modes. The Fisher information matrix in this scenario shows a phase-dependent profile $\mathcal{F}(\Delta \phi_1, \Delta \phi_2)$, meaning that without adaptive strategies the asymptotic precision will be different depending on the actual phase values. In particular, by looking at the inverse of $\mathcal{F}$, we obtain $\min_{\Delta \phi_1, \Delta \phi_2} \mathrm{Tr}(\mathcal{F}^{-1})  \simeq 3.866$, which is obtained for six different phase pairs $(\tilde{\Delta \phi_1}, \tilde{\Delta \phi_2})$. For those pairs, minimum asymptotic quadratic loss is achieved.

Bayesian protocols require in general expensive computational resources, due to the need of evaluating complex integrals to determine the normalization constant $\mathcal{N}$, as well as the estimated values and their corresponding covariance matrices. A possible solution is to perform a discretization of the parameters space, thus converting integrals to sums. In this case, the bin size has to be chosen depending on the minimum error expected at the end of the estimation process. However, such solution becomes quickly unmanageable when the number of parameter increases, since such discretization has to be performed in a $n$-dimensional space. A different solution has been explored in \cite{granade2012online} for Bayesian learning problems by using a Sequential Monte Carlo (SMC) approach. Indeed, Monte Carlo methods seems to be a natural solution, due to their capability of reaching convergence independently from the integration space dimension. The SMC method approximates the infinite dimensional support $\bm{\phi}$ with a finite number $M$ of elements $\bm{\phi}_{i}$, called particles, with associated probability weights $w_i$. The error in the approximation can be arbitrarily reduced by increasing the number of particles, leading to a trade-off between computational time and accuracy of the approximation. In the context of Bayesian analysis, any distribution $\tilde{p}(\bm{\phi})$ in the particles approximation is expressed as $\tilde{p}(\bm{\phi}) \approx \sum_{i=1}^{M} w_i \delta(\bm{\phi}-\bm{\phi}_i)$.

We now consider the case of an initial prior knowledge $p(\bm{\phi})$ corresponding to a uniform distribution. In the particles scenario, this prior information is approximated by a set of $M$ randomly drawn pair of phases $\bm{\phi}_i$ with equal weights $w_i=1/M$ to satisfy the normalization condition $(\sum_{i=1}^{M} w_i = 1)$. During the experiment, the information about the unknown phases $\bm{\phi}$ is updated according to the Bayes rule after each measurement outcome $d$. In the particle approximation, having fixed control phases, this corresponds to updating the particle weights as  $w_i \rightarrow w_i \, p(d \vert \bm{\phi}_i,\bm{\Phi})$, while keeping the particles $\{\bm{\phi}_{i}\}$ unchanged. The estimation of $\bm{\phi}$ is then provided by the expectation value of the posterior distribution $\hat{\bm{\phi}} = \int_{}^{} d\bm{\phi} \, \bm{\phi} \, p(\bm{\phi} \vert d,\bm{\Phi}) \approx \sum_{i=1}^{M} w_i \bm{\phi}_i$. As discussed in \cite{granade2012online}, the particle approximation needs some additional steps to avoid the introduction of further errors throughout the estimation process. In particular, after a few iterations the non-zero weights will be mostly concentrated on a small subset of $\{\bm{\phi}_{i}\}$, reducing the validity of the approximation. To avoid such effect, it is possible to employ resampling techniques \cite{liuwest}. More specifically, when the particle weights become too concentrated according to a given threshold condition, a new set of particles $\{ \bm{\phi}'_i \}$ is generated by adding a small random perturbation to the original particles (see Supplementary information IA for more details). The weights are then reset to $w'_i = 1/M$, and the estimation process restarts. Within this framework, we now have to define the adaptive rule to determine the value of the control parameters at each step depending on the actual knowledge. More specifically, at each step of the estimation process one has to decide the control parameters $\bm{c}$ (here, the additional phases $ \bm{\Phi}$) for the next probe. To this end, we consider different strategies. 

A first approach (i) is based on choosing the control phases according to $\bm{\hat{\phi}} + \bm{\Phi} \simeq \bm{\delta \phi}$, where $\bm{\delta \phi} = \mathrm{argmin}_{\Delta \phi_1, \Delta \phi_2} \mathrm{Tr}(\mathcal{F}^{-1})$. This strategy  looks to set the interferometer phases $\bm{\Delta \phi}$ to those values leading to a minimum bound for $L(\bm{\phi}, \bm{\hat{\phi}})$ according to the Cram\'er-Rao inequality. While this approach is tailored to work in the asymptotic regime of large $N$, its performances are not guaranteed to be optimal for small $N$. An upside of this approach is that setting the control parameters does not require complex optimization steps, since an analytic rule can be easily defined. 

In order to devise a strategy working in the small $N$ regime, one can consider a second strategy (ii) which is specifically tailored to work for all values of $N$. To this end, we adapted the protocol described in \cite{granade2012online} to the multiparameter scenario implemented by our system. By this approach, the choice of the control phases is performed to optimize a given figure of merit, known as utility function ($U$). Canonical choices for $U$ are information gain or quadratic loss. In our case, we choose $U(\bm{\hat{\phi}}) = \mathrm{Tr}[\mathrm{Cov}(\bm{\hat{\phi}})]$. Hence, at each step the minimization algorithm finds the best control phases $\bm{\Phi}$ that, averaged over all possible measurement outcomes, leads to a minimum value for the sum of the parameters confidence intervals. This is thoroughly discussed in Sec. IB of the Supplementary Information. Given that this method relies on numerical optimization steps, it is more expensive in terms of computational resources than the previous strategy based on the Fisher information matrix. Conversely, it provides the advantage of searching the optimal control phases for all values of $N$, thus covering the limited data regime where asymptotic approaches may not be the proper choice.

\begin{figure}[ht!]
\centering
\includegraphics[width=.49\textwidth]{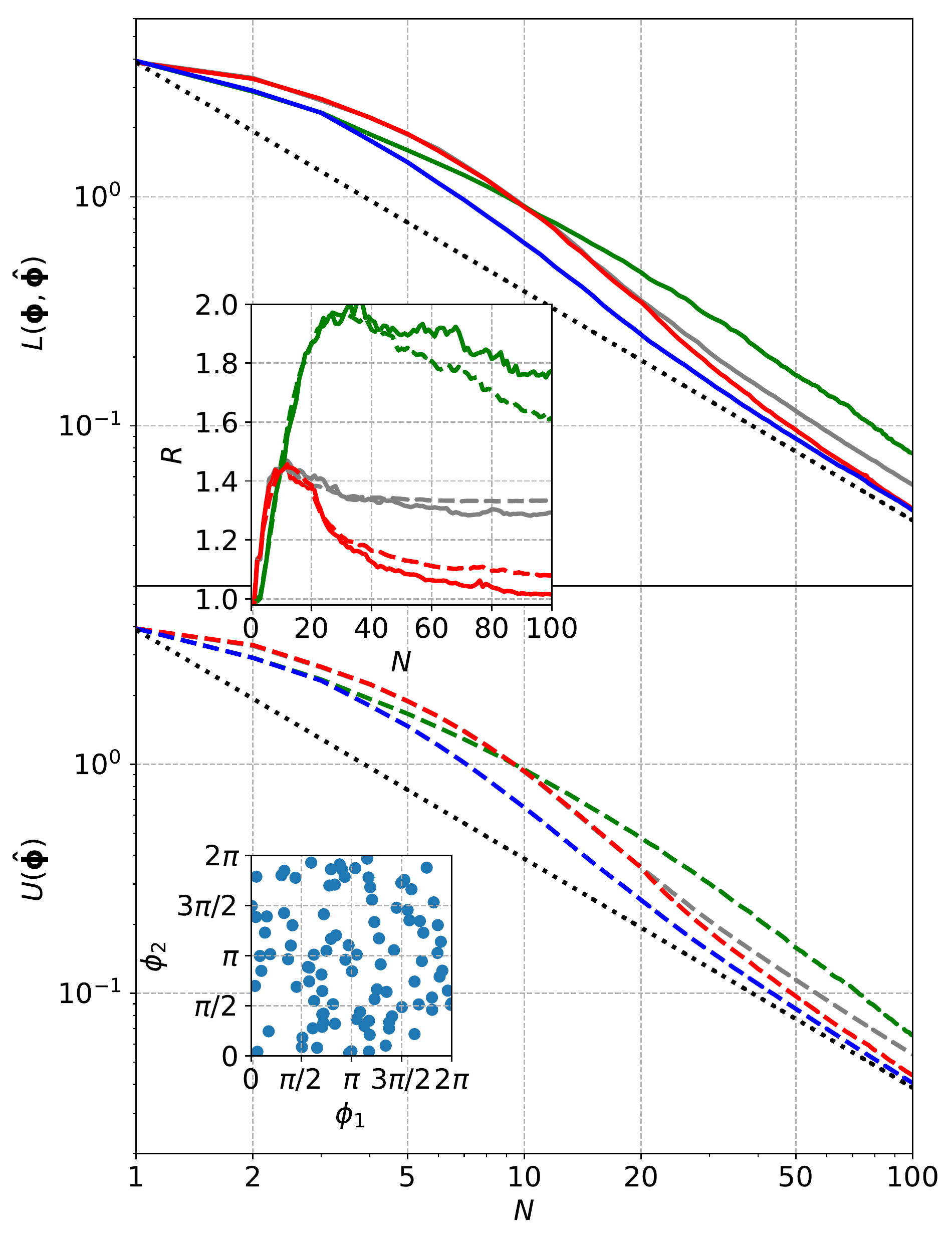}
\caption{Numerical simulations of Bayesian adaptive protocols. For $N_{\mathrm{ph}} = 100$ pair of phases, we simulated the performance of the different strategies described in the main text, by averaging for each phase over $N_{\mathrm{exp}}=100$ different runs. Top: quadratic loss $L(\bm{\phi}, \bm{\hat{\phi}})$ (solid lines). Bottom: utility function $U(\bm{\hat{\phi}}) = \mathrm{Tr}[\mathrm{Cov}(\bm{\hat{\phi}})]$ (dashed lines), corresponding to the sum of the parameters confidence intervals.
Inset: (top) ratio $R$ between the performances of each protocol, compared with the optimized strategy (ii). $R$ is computed both for $L(\bm{\phi}, \bm{\hat{\phi}})$ (solid lines) and $U(\bm{\hat{\phi}})$ (dashed lines), referring the same colors of the main panels. (bottom) two-dimensional map of uniform-distributed couples of phases drawn for the simulations. Green lines: approach (i) based on the Fisher information matrix. Red lines: approach (i') which includes first $N=20$ events with random control parameters, while for $N>20$ works as (i). Blue lines: optimized approach (ii). Grey lines: benchmark approach with random control parameters (iii). Dotted black lines: Cram\'er-Rao bound for the asymptotic regime.}
\label{fig:simulations} 
\end{figure}

We have then performed numerical simulations to characterize the performances of the two algorithms. More specifically, we have sampled $N_{\mathrm{ph}} = 100$ random pairs of phases $(\phi_1, \phi_2)$ in the interval $[0, 2\pi]\times[0, 2\pi]$. For each pair, we simulated $N_{\mathrm{exp}} = 100$ estimation processes where $N = 100$ single-photon probes are sent in the interferometer. The results are shown in Fig. \ref{fig:simulations}. We first tested the performances of both algorithms (i) and (ii). We observe that, concerning strategy (i), the protocol fails to approach the Cram\'er-Rao bound even for $N \sim 100$. This is related to the periodicity of the likelihood function, which presents multiple equivalent points. Approach (i) seeks for setting the phase differences $\bm{\Delta \phi}$ to a fixed point, and it is not able to resolve such periodicity issue. Better results are obtained by applying at each step a random (but known) set of control phases (iii), which shows better convergence while not reaching the Cram\'er-Rao bound. However, the application of this strategy is capable of resolving the multiple periodicity. One can then consider a modified version (i') of the asymptotic protocol (i), where the first $K$ control phases are drawn from a uniform distribution, while for $N>K$ the strategy works as (i). Numerical evidence shows that the best choice for this parameter is $K \sim 20$. We observe that, with this modified strategy, the Cram\'er-Rao bound is approached for $N \sim 50$. Better results are obtained with the optimized strategy (ii), in particular in the small $N$ regime. For $N > 60$, we observe that both strategies (i') and (ii) provide similar performances since the experiment progressively approaches a large $N$ scenario where the Fisher information matrix defines the system sensitivity. In this work we experimentally implement the optimal strategy to guarantee a faster convergence of the estimation process.

\subsection*{Integrated circuit for multiphase estimation}

\begin{figure}[ht!]
\centering
\includegraphics[width=.49\textwidth]{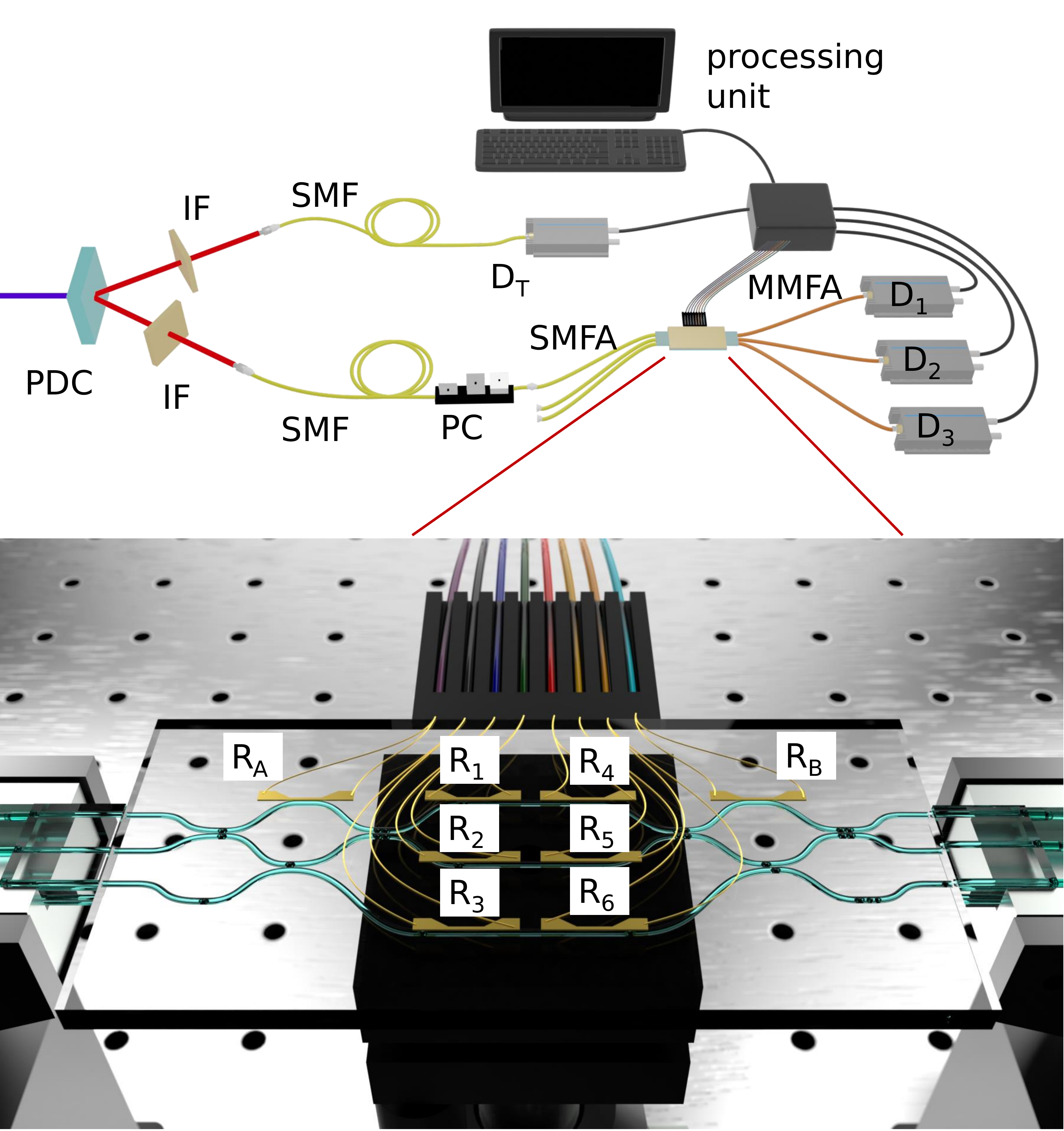}
\caption{Experimental platform. A type-II parametric down conversion source (PDC) generates photon pairs, which are spectrally selected via interference filters (IF) and coupled to single-mode fibers (SMF). One of the photons is directly measured by detector $D_{T}$ acting as trigger for the experiment. The other photon, after polarization compensation (PC), can be injected in any of the three input ports of the interferometer via a single-mode fiber array (SMFA). After evolution, photons are collected via a multi-mode fiber array (MMFA) and measured through detectors $D_{i}$, with $i=1,2,3$. Coincidences between $D_{T}$ and any of $D_{i}$ are recorded via a time-to-digital converter. The results of the measurement are processed and employed to apply the adaptive protocols. The layout of the integrated circuit (shown in the bottom panel) includes 8 resistors to modulate the input transformation ($R_{A}$), the output one ($R_{B}$), and the internal phases ($R_{i}$, with $i=1,\ldots,6$) as described in the main text.}
\label{fig:setup} 
\end{figure}

The platform employed in this experiment is an integrated three-arm interferometer. This system has been employed in Ref. \cite{polino2019experimental} for the simultaneous estimation of two relative phase shifts $\bm{ \phi}= (\phi_1, \phi_2)$ between the arms of a three mode interferometer (Fig. \ref{fig:setup}). We first discuss the circuit layout and parameters, while we subsequently describe the working condition used for the multiphase estimation experiments reported below.

The platform is a three-arm interferometer realized in a glass chip through femtosecond laser writing \cite{della2008micromachining, gattass2008femtosecond}.  The interferometer, optimized for operation at $\lambda = 785$ nm, is implemented by two cascaded tritters (three-mode  beam splitters) $A$ and $B$ interspersed with phase shifters. Each tritter is decomposed in a 2-D planar configuration \cite{reck1994experimental} consisting of three balanced directional couplers and one phase shifter $\phi_\mathrm{T}^A$ ($\phi_\mathrm{T}^B$) for tritter $A$ ($B$). These phase shifters, as well as those placed between the two tritters, can be tuned by means of the thermo-optic effect, using microresistors that are patterned in a thin gold layer covering the chip surface. When an electrical current is applied to the resistor, an optical path change on the waveguide is induced by the dissipated heat \cite{flamini2015thermally}. In particular, let us consider the dissipated power $P_i=R_{i}\,I_{R_i}^{2}$ on resistor $R_i$ subjected to a current $I_{R_i}$, where we also include that the value of the resistor depends on the current due to its temperature change. The two induced relative phase shifts $\bm{\Delta\phi}=(\Delta\phi_{1}, \Delta\phi_{2})$  between the arms of the interferometer with respect to the reference mode, have the following general dependence on the dissipated powers: 
\begin{equation}
\label{phases}
\Delta\phi_{j}=\phi_{j0}+\sum_{i=1}^6 \left(\alpha_{ji}P_{i}+\sum_{k=i}^6 \alpha_{jik}^{NL}P_{i} P_{k} \right),
\end{equation}
where $j=1,2$ and $\phi_{j0}$ stands for the static phases of the interferometer. Parameters $\alpha_{ji}$ and $\alpha_{j,i=k}^{NL}$ are the linear and quadratic response coefficients relative to the dissipated power $P_{i}$, respectively, while $\alpha_{j,i\ne k}^{NL}$ represent the nonlinear coefficients associated to the product of the two powers $P_{i}$ and $P_{k}$ to include cross-talk effects. In our device 8 independent resistors are present (Fig. \ref{fig:setup}). Resistors $R_A$ and $R_B$ are exploited to tune tritter phases $\phi_\mathrm{T}^A$ and $\phi_\mathrm{T}^B$, respectively. Conversely, resistors $R_1$, $R_4$ along mode 1, $R_2$, $R_5$ along mode 2 and $R_3$, $R_6$ along mode 3, are employed to tune the internal relative phase shifts of the interferometer, according to \eqref{phases}. The operations of tritters $A$ and $B$ are described through the unitary evolutions $U_A$ and $U_B$, respectively, while the action of each phase shifter along mode $i$ is described through a unitary matrix $PS_i$ ($i=1,2,3$). The overall evolution $U^{\mathrm{tot}}$ of the interferometer is given by $U^{\mathrm{tot}}=U_B (\prod_{i=1}^{3} PS_i) U_A$.

In order to characterize the relevant parameters necessary to fully describe the evolution of the interferometer, we measure the output probabilities when single photons are injected along input 1, tuning the current applied on each resistor. The probabilities have been theoretically modeled by modifying the ideal expression with additional terms, taking into account non-ideal visibilities and dark counts of the detectors. In this way, we performed an overall fit of all the measured probabilities to determine the 58 chip parameters (Supplementary Information II) and finely reconstruct the likelihood probability $p(d \vert \bm{\Delta\phi})$ of our system. 

According to the scheme of Fig. \ref{fig:concept}b, the unknown phases to be estimated are the pairs ($\phi_1, \phi_2$), relative to the chosen reference arm $\phi_{\mathrm{ref}}$. The 8 resistors allow us to finely tune and control all the relevant phase shifts of the interferometer. The tritters phases can be tuned and are chosen in order to maximize the sensitivity of the interferometer. Using single photon probes, the optimal configuration for our interferometer employs mode 1 as input and mode 2 as reference. In this case, the trace of the inverse of Fisher Information matrix, minimized over all possible internal phases, is $\mathrm{Tr}[(\mathcal{F}_{\mathrm{exp}})^{-1}] = 4.2$. The unknown phases $\bm{\phi}=(\phi_1, \phi_2)$ are tuned by means of resistors $R_4, R_5$ and $R_6$, according to \eqref{phases}, while the control phases $\bm{\Phi}=(\Phi_1, \Phi_2)$ are tuned by resistors $R_1$ and $R_2$ (see Methods).

\subsection*{Experimental adaptive multiphase estimation}

We perform the experiment by continuously adapting the present tunable circuit following the optimized Bayesian-SMC method [strategy (ii)]. This allows us to achieve best attainable estimation with a limited number of resources. The probes are heralded single photons at $785$ nm generated by a degenerate type-II SPDC process inside a BBO crystal, pumped by a pulsed $392.5$ nm laser. A photon from each pair is sent through the circuit, entering in input 1, and acts as probe, while the other photon acts as the trigger for the heralding process (see Fig. \ref{fig:setup}). An event is then recorded as the coincidence between the trigger detector and one of the three outputs of the circuit. The interaction of the probe with the chip operator encodes information about $\bm{\phi}$ onto its state. Finally, the result of the measurement is collected and used to identify the optimal settings for the next experimental step. 

\begin{figure}[ht!]
\centering
\includegraphics[width=.49\textwidth]{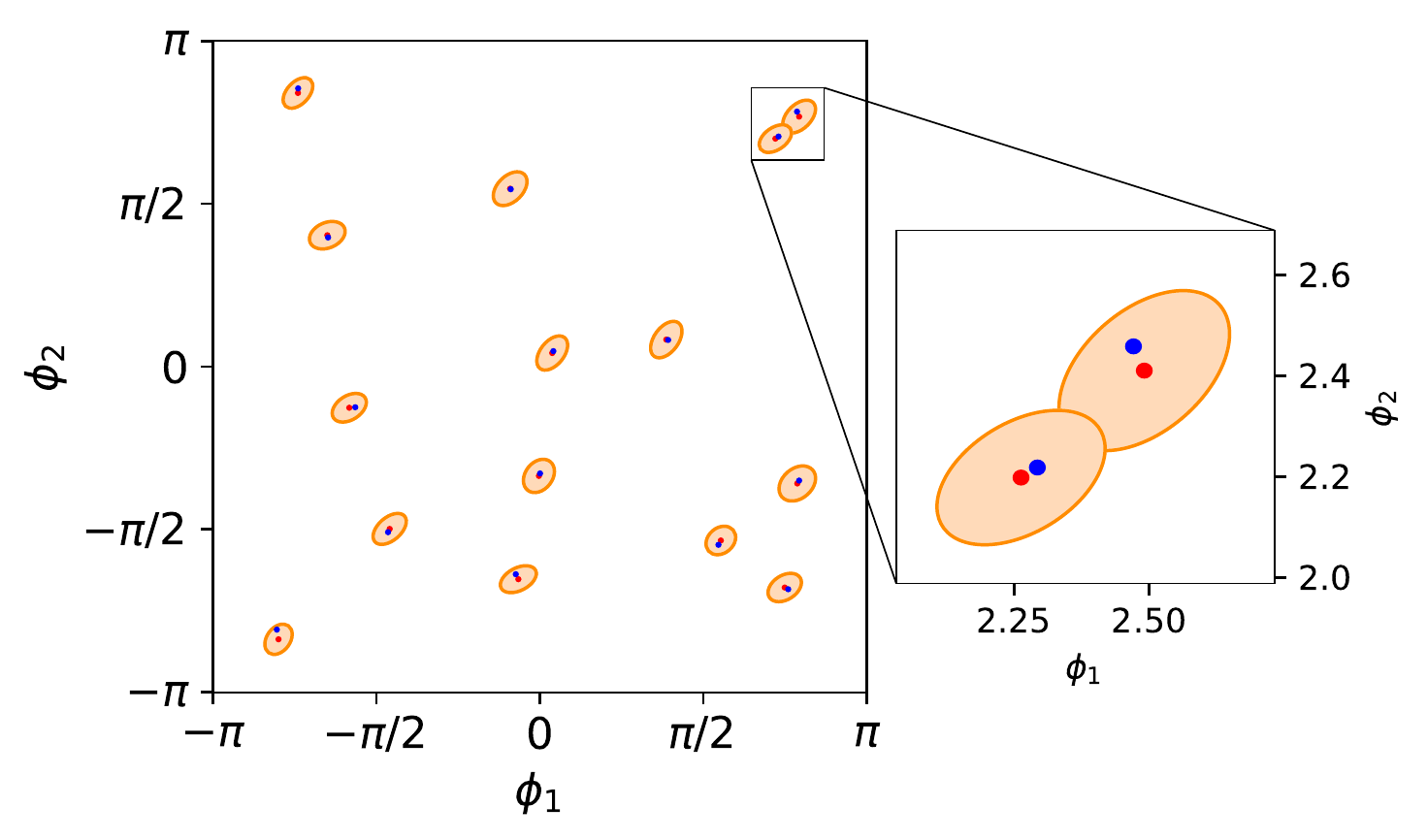}
\caption{Experimental simultaneous estimations of $N_{\mathrm{ph}} = 15$ different uniform-distributed pairs of phases. The estimation process uses an amount of $N = 100$ resources and Bayesian adaptive approach. Dark orange regions represent the error in the estimation obtained from the covariance matrix  Each estimated pair (red dot) is distant from true set value (blue dot) within the error (orange area), thus confirming the good performance of the algorithm. 
}
\label{fig:fig5} 
\end{figure}

\begin{figure*}[ht!]
\centering
\includegraphics[width=1.\textwidth]{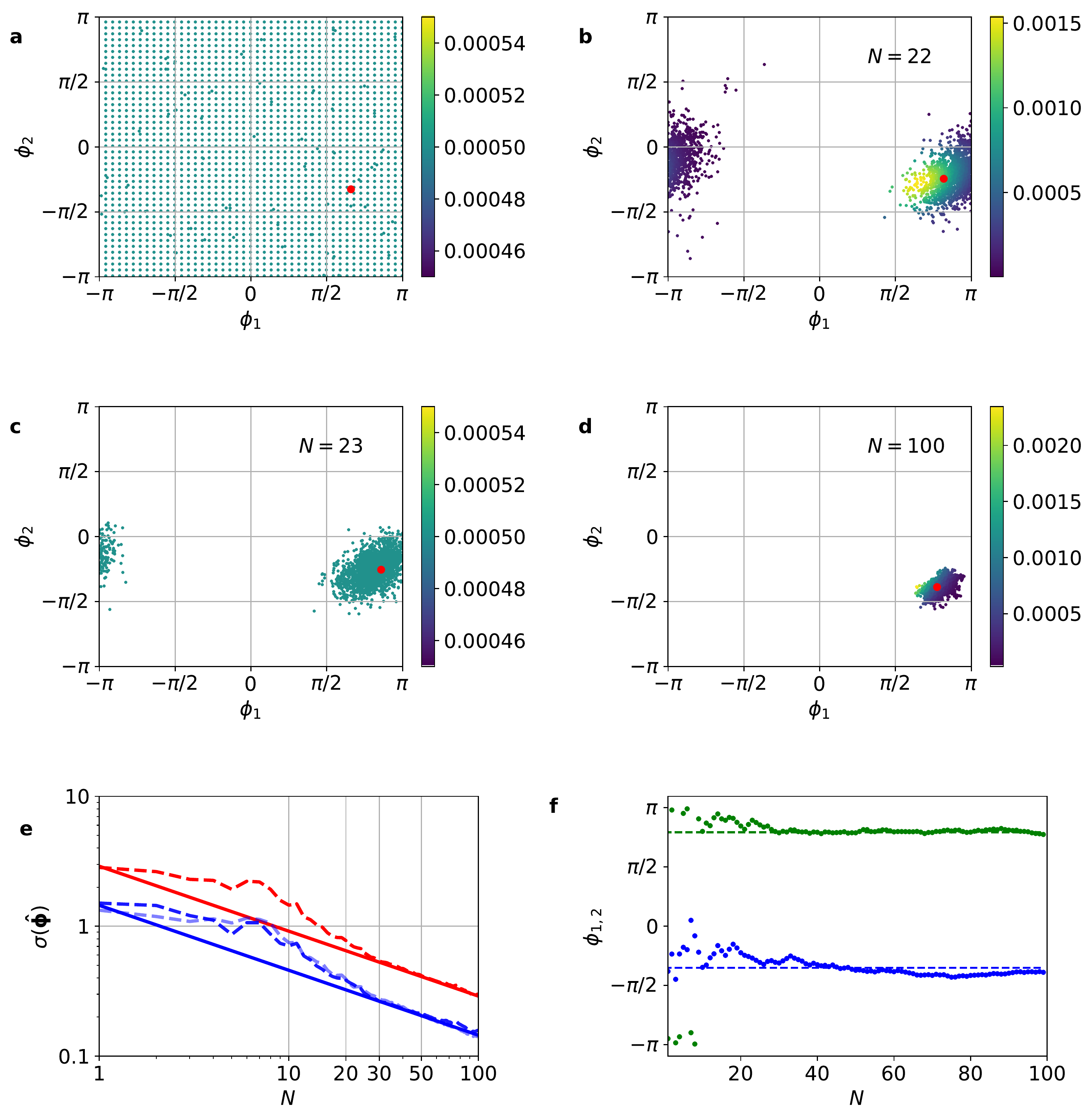}
\caption{Example of experiment for the estimation of two phases (red dots). {\bf a}, A uniform grid is generated as initial support for the prior distribution. {\bf b-d}, Evolution of the posterior distribution during experiment for three subsequent moments. In particular, the distribution before and after the resampling are shown respectively in {\bf b} and {\bf c}, where particles are rearranged in order to eliminate zero weight cases. The new posterior weights are uniform, while particles are distributed closer to the estimated phases. {\bf e}, Study of standard deviation in estimation of the single phases (blue solid lines) and their sum (red solid lines). The saturation of their CRB (dashed lines) occur for small $N$. {\bf f}, Experimental estimated pair of phases as function of the number $N$ of adopted probes (dots). Dashed lines indicates true set values of the phases.}
\label{fig:fig4} 
\end{figure*}

\begin{figure*}[ht!]
\centering
\includegraphics[width=1.\textwidth]{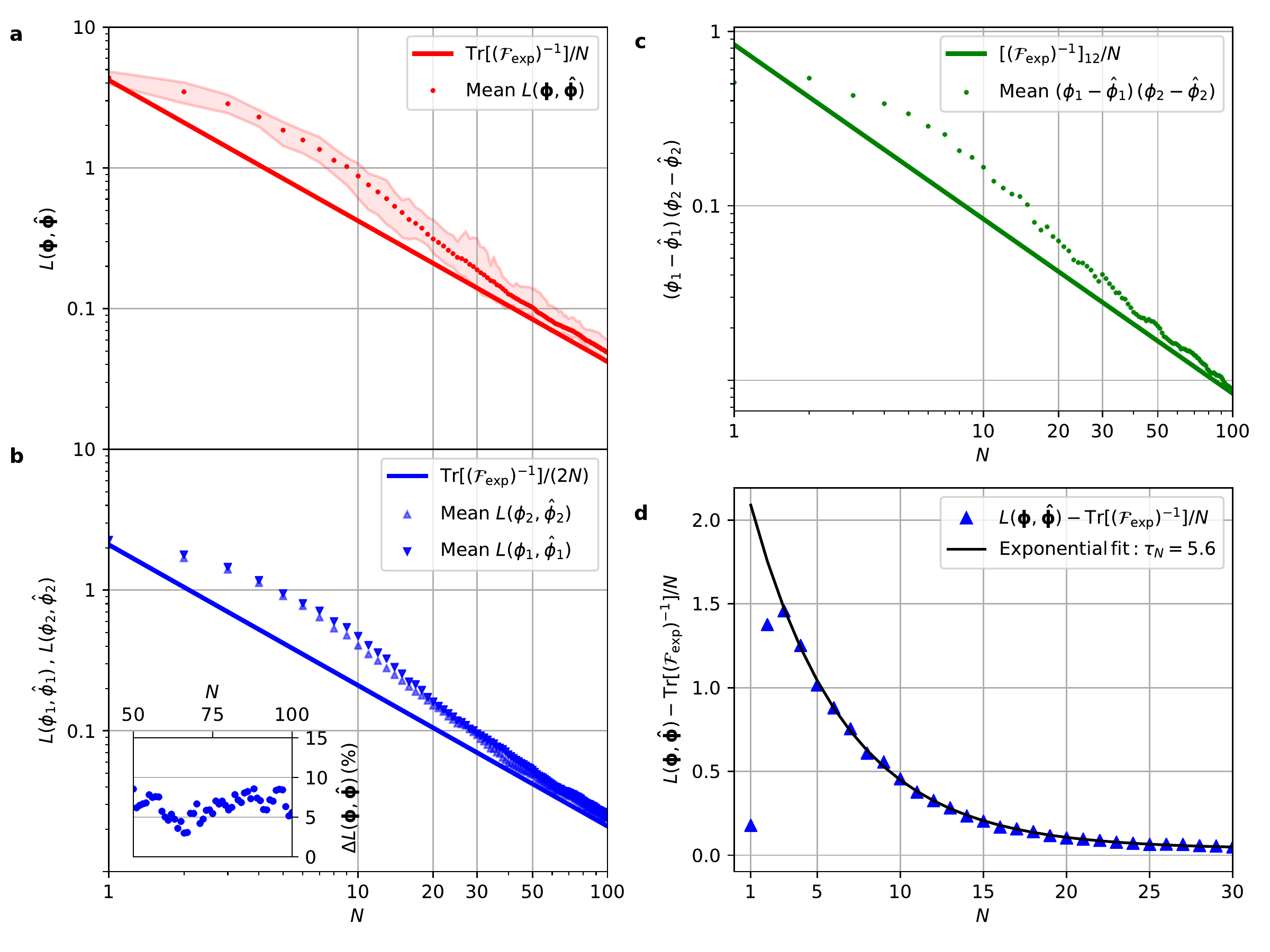}
\caption{Experimental simultaneous estimations of $N_{\mathrm{ph}} = 15$ different pairs of phases using Bayesian adaptive protocol. Quadratic loss is averaged for each phase over $N_{\mathrm{exp}} = 100$ independent runs. {\bf a}, Comparison between overall quadratic loss $L(\bm{\phi}, \bm{\hat{\phi}})$ (red dots) and $\mathrm{Tr}[(\mathcal{F}_{\mathrm{exp}})^{-1}]/N$ (red solid line). The performances are in agreement with the numerical simulations. Red shaded regions represents the interval where $L(\bm{\phi}, \bm{\hat{\phi}})$ relative to each single phase (averaged on its $N_{\mathrm{exp}}$ runs) can be found. {\bf b}, Analysis of diagonal elements of CRB by comparing quadratic loss relative to the single phase of the estimated pair $L(\phi_i, \hat{\phi_i})$ (with $i=1,2$) (blue triangles) and $\mathrm{Tr}[(\mathcal{F}_{\mathrm{exp}})^{-1}]/(2N)$ (blue solid line). The algorithm shows symmetric optimal performances for estimation of both parameters, by using the same amount of resources. This feature is highlighted by the inset panel, where the ratio $\Delta L(\bm{\phi}, \bm{\hat{\phi}})$ between the difference of the two estimations and the bound value is reported. {\bf c}, Analysis of phase correlations by comparing off-diagonal terms of $\mathcal{F}_{\mathrm{exp}}^{-1}$ (see Sec. III of Supplementary Information) and $[(\mathcal{F}_{\mathrm{exp}})^{-1}]_{12}/N$ (green solid line). {\bf d}, Estimation of  convergence time ($\tau_N$) to CRB. The value can be estimated by fitting the distance between the averaged $L(\bm{\phi}, \bm{\hat{\phi}})$ and CRB, after $N>2$. The adopted fit function is $a+b \exp{(-N/\tau_N)}$, with $a,b,\tau_N \in \mathbb{R}$ the fitting parameters, leading to $\tau_N=5.6$. The choice of this function is performed to provide a reasonable estimation of $\tau_N$, as the number of probes necessary to approach the CRB.}
\label{fig:fig6} 
\end{figure*}

The phases $\bm{\phi}$ to be estimated can be chosen by setting the currents flowing in three resistors $R_4, R_5, R_6$ (see Methods). In order to test the protocol over different estimation experiments, we have identified $N_{\mathrm{ph}}=15$ pair of phases uniformly distributed (Fig. \ref{fig:fig5}). Resistors $R_1, R_2$ are used to tune the control phases necessary for the adaptive strategy. After the first event, where currents $I_{R_1},I_{R_2}$ are chosen at random, we implement strategy (ii): optimal control phases $\bm{\Phi}$ are calculated by minimizing the expected posterior variance. The nearest available control currents $I_{R_1},I_{R_2}$, limited by the precision of our power supply (Keithley 2230), are calculated and effective control phases are applied to the device. The calculation of the prior distribution for each step is made through the particle approximation. A uniform grid of $n=2000$ pairs of phases (Fig. \ref{fig:fig4}a) is assumed as initial set for the prior distribution. This choice is performed yo avoid any possible harmful periodicity during the estimation process. Examples of prior information evolution during an experiment are reported in Fig. \ref{fig:fig4}b-d. In Fig. \ref{fig:fig4} c the resampling step is shown, where particles with zero weight of the previous step (Fig. \ref{fig:fig4}b) are rearranged in more significant locations (see Sec. IA of Supplementary Information for more details). Each pair is estimated $N_{\mathrm{exp}}=100$ times, adopting $N=100$ resources (photons) as for the numerical simulations discussed above. Single experiments are reported in details in Sec. II of Supplementary information. Algorithm performances are shown in Fig. \ref{fig:fig6}. A first evaluation consists in averaging the experimental quadratic loss for each pair of phases over all $N_{\mathrm{exp}}$ independent runs. As a result, the overall quadratic loss $L(\bm{\phi}, \bm{\hat{\phi}})$ saturates the CRB with a limited number of resources, in agreement with the numerical simulations described above. Furthermore, saturation occurs  both for off- and diagonal matrix elements of the CRB. In particular, the latter show that the CRB is reached with similar performances in the estimation of both phases. This result is a fundamental feature for multiparameter metrology tasks when both parameters are treated equally. In our case the resulting difference in estimation of the two parameters is less than $10\%$, when compared to the sensitivity bound. Furthermore, a heuristic estimation of the convergence time to saturate CRB can be calculate by studying the difference $L(\bm{\phi},\bm{\hat{\phi}})-\mathrm{Tr}[(\mathcal{F}_{\mathrm{exp}})^{-1}]/N$. A characteristic time can be computed by using  $a+b \exp{-(N/\tau_N)}$ as fit function, with $a,b,\tau_N \in \mathbb{R}$ the fitting parameters. The value obtained for $\tau_N$ is $\tau^{\mathrm{fit}}_N=5.6$, which underlines the good performance of the adaptive adopted technique in using small number of probes. Another significant property of Bayesian approach is the ability to provide the statistical error in each step of the estimation process, calculated as the variance of the posterior distribution. Final estimated pairs fall on average within the error from true set values of phases (Fig. \ref{fig:fig5}). All these experimental results demonstrate the quality of Bayesian-SMC strategy, confirming it as largely suitable for multiparameter estimation problems. Implementation of this strategy has been enabled onlyby the high reconfigurability of our employed integrated device, which highlights the fundamental role of an appropriate platform for metrology tasks which involve more than one parameter.        

\section*{Discussion}
Multiparameter estimation is a fundamental problem for the realization of realistic quantum sensors in several scenarios. In this task, there are still several open problems and a comprehensive framework has yet to be defined. Hence, it is crucial to identify an experimental platform versatile enough to address different possible approaches. Multiphase estimation provides an ideal scenario with different practical applications. Furthermore, it represents a testbed for different multiparameter estimation protocols. Applying these to real world scenario requires a further step, that is, the optimization of the available resources, so as to attain the minimum reachable uncertainties after a sufficiently small number of measurements. This can be achieved by implementing adaptive strategies.

Here, we have reported the first experimental implementation of a multiphase Bayesian adaptive protocol on an integrated platform, optimized to operate in the limited data regime. We have reviewed different adaptive strategies and selected the one optimizing the cost function given by the trace of the covariance matrix. This has been employed to perform several simultaneous estimations of uniformly distributed pairs of phases. As we have shown, the achievable bounds are attained for both unknown phases after a limited number of $N\sim 40$ probes. Our experiment permits to underline the suitability of such an integrated circuit for performing multiparameter estimation tasks, as well as to exploit the capabilities of the proposed Bayesian adaptive strategy.  

This work provides a versatile approach for future perspectives in multiparameter quantum metrology. In particular, these techniques can be directly generalized for multi-photon quantum probes which would provide insight on the achievable quantum accuracy limit. At the same time, the algorithm here described can be applied to more complex integrated platforms, which enable optimized extraction of information. Further perspectives include the study of different multiparameter scenarios, as well as practical applications to quantum sensing of delicate samples \cite{crespi2012measuring}. 

\section*{Methods}
\subsection*{Tuning of circuit parameters for adaptive two-phase estimation}

We discuss in more details how we exploit the phases in our interferometer. The pair ($\phi_1, \phi_2$) represents the unknown phases relative to a reference arm with phase  $\phi_{\mathrm{ref}}$ (Fig. \ref{fig:concept}b).
All the relevant phases of the  circuit can be finely tuned by means of 8 resistors.

The first step performed aimed at finding the optimal choice for the tritter phases $\phi_\mathrm{T}^A$, $\phi_\mathrm{T}^B$ to maximize the sensitivity of the interferometer. To achieve this goal, we first evaluate the Fisher information matrix $\mathcal{F}_{\mathrm{exp}}$ associated to the device from the experimentally estimated parameters,  Then, we numerically minimize $\mathrm{Tr}[(\mathcal{F}_{\mathrm{exp}})^{-1}]$  over all possible values of $\phi_\mathrm{T}^A$, $\phi_\mathrm{T}^B$ and internal phases $\bm{\Delta \phi}$, in the allowed range of dissipated powers (the upper threshold being $\sim 1$ W, to avoid possible damages to the resistors). We identify such minimum for all combinations of possible inputs and reference arms. The best scenario for our interferometer corresponds to use mode 2 as reference mode and arm 1 as input mode for single photons, with the following values of phases: $\phi_\mathrm{T}^A=1.49$ rad, $\phi_\mathrm{T}^B=0.72$ rad, $\Delta \phi_1= -3.07$ rad and $\Delta \phi_2 = 0.34$ rad. In this working point, the trace of the inverse of Fisher Information matrix is $\mathrm{Tr}[(\mathcal{F}_{\mathrm{exp}})^{-1}] = 4.2$. We now have to assign each resistor $R_i$ ($i=1, \ldots, 6$) to tune both the unknown phase shifts $\bm{\phi}=(\phi_1, \phi_2)$, and the control phases $\bm{\Phi}=(\Phi_1, \Phi_2)$ for the adaptive algorithms. More specifically, we choose to employ resistors $R_4, R_5$ and $R_6$ to tune $\bm{\phi}$. Conversely, the control phases $\bm{\Phi}$ are those modified by dissipating power in $R_1$ and $R_2$. Hence, considering \eqref{phases} as $\bm{\Delta\phi}=\bm{\phi}+\bm{\Phi}$, we find the following expressions:
\begin{eqnarray}
\phi_{j}&=&\phi_{j0}+\sum_{i=4}^6 \left(\alpha_{ji}P_{i}+\sum_{k=i}^6 \alpha_{jik}^{NL}P_{i} P_{k} \right)
\\
\Phi_{j}&=&\sum_{i=1}^2 \left(\alpha_{ji}P_{i}+\sum_{k=i}^2 \alpha_{jik}^{NL}P_{i} P_{k} \right)\;,
\end{eqnarray}
with $j=1,2$. Note that, in principle, only 4 resistors would be sufficient to tune independently the 4 phase shifts (2 unknown and 2 controls). However, we employed 5 resistors in order to obtain large tunability of the device within limits of the damage threshold of each resistor.

\providecommand{\noopsort}[1]{}\providecommand{\singleletter}[1]{#1}%

\section*{Acknowledgments}
We acknowledge very fruitful discussions with Nathan Wiebe, and useful discussions with Francesco Hoch on the integrated device calibration. This work is supported by the Amaldi Research Center funded by the  Ministero dell'Istruzione dell'Universit\`a e della Ricerca (Ministry of Education, University and Research) program ``Dipartimento di Eccellenza'' (CUP:B81I18001170001), by MIUR via PRIN 2017 (Progetto di Ricerca di Interesse Nazionale): project QUSHIP (2017SRNBRK), by QUANTERA HiPhoP (High dimensional quantum Photonic Platform; grant agreement no. 731473), and by the Regione Lazio programme “Progetti di Gruppi di ricerca” legge Regionale n. 13/2008 (SINFONIA project, prot. n. 85-2017-15200) via LazioInnova spa.





\end{document}


\title{Supplementary material: Experimental adaptive Bayesian estimation of multiple phases with limited data} 

\affiliation{Dipartimento di Fisica - Sapienza Universit\`{a} di Roma, P.le Aldo Moro 5, I-00185 Roma, Italy}

\author{Mauro Valeri}
\affiliation{Dipartimento di Fisica, Sapienza Universit\`{a} di Roma, Piazzale Aldo Moro 5, I-00185 Roma, Italy}
\author{Emanuele Polino}
\affiliation{Dipartimento di Fisica, Sapienza Universit\`{a} di Roma, Piazzale Aldo Moro 5, I-00185 Roma, Italy}
\author{Davide Poderini}
\affiliation{Dipartimento di Fisica, Sapienza Universit\`{a} di Roma, Piazzale Aldo Moro 5, I-00185 Roma, Italy}
\author{Ilaria Gianani}
\affiliation{Dipartimento di Fisica, Sapienza Universit\`{a} di Roma, Piazzale Aldo Moro 5, I-00185 Roma, Italy}
\affiliation{Dipartimento di Scienze, Universit\`a degli Studi Roma Tre, Via della Vasca Navale 84, 00146, Rome, Italy}
\author{Giacomo Corrielli}
\affiliation{Istituto di Fotonica e Nanotecnologie, Consiglio Nazionale delle Ricerche (IFN-CNR), Piazza Leonardo da Vinci, 32, I-20133 Milano, Italy}
\affiliation{Dipartimento di Fisica, Politecnico di Milano, Piazza Leonardo da Vinci, 32, I-20133 Milano, Italy}
\author{Andrea Crespi}
\affiliation{Istituto di Fotonica e Nanotecnologie, Consiglio Nazionale delle Ricerche (IFN-CNR), Piazza Leonardo da Vinci, 32, I-20133 Milano, Italy}
\affiliation{Dipartimento di Fisica, Politecnico di Milano, Piazza Leonardo da Vinci, 32, I-20133 Milano, Italy}
\author{Roberto Osellame}
\affiliation{Istituto di Fotonica e Nanotecnologie, Consiglio Nazionale delle Ricerche (IFN-CNR), Piazza Leonardo da Vinci, 32, I-20133 Milano, Italy}
\affiliation{Dipartimento di Fisica, Politecnico di Milano, Piazza Leonardo da Vinci, 32, I-20133 Milano, Italy}
\author{Nicol\`o Spagnolo}
\affiliation{Dipartimento di Fisica, Sapienza Universit\`{a} di Roma, Piazzale Aldo Moro 5, I-00185 Roma, Italy}
\author{Fabio Sciarrino}
\email{fabio.sciarrino@uniroma1.it}
\affiliation{Dipartimento di Fisica, Sapienza Universit\`{a} di Roma, Piazzale Aldo Moro 5, I-00185 Roma, Italy}

\maketitle

\section{Adaptive Bayesian estimation based on Sequential Monte Carlo approach}

As explained in the main text, the machine learning technique exploited to realize two-phase estimation experiments is based on approximating the prior probability distribution support with $M$ discrete particles \cite{granade2012robust}. More specifically, a probability weight $w_i$ is associated to each $i-$particle by keeping normalization $\sum_{i=1}^{M} w_i = 1$. Then, the posterior distribution is updated according to the Bayes rule. In this section we discuss two aspects of this technique: the resampling strategy, and the utility function chosen to tune the control parameters. Finally, we report some experimental data to provide an overview of the overall estimation process.

\subsection{Resampling strategy}
Fig. 5a of the main text shows that the initial support is uniformly covered by particles. During the estimation protocol, the updating process according to the Bayes rule modifies the particles weights towards more likely phases values. Greater weights are attribute to particles closer to the estimated values. Conversely, distant particles tend to zero weights according to the normalization rule (Fig. 5b of the main text), thus bringing no useful information. Furthermore, the estimation sensitivity of the unknown phase pairs is limited by the initial density of particles around the true values. To solve these problems, resampling technique can be adopted to update particles in more significant positions. Following Ref. \cite{granade2012robust}, the employed resampling technique is based on introducing, at particular steps of the process, perturbations on the particles (\{$\bm{\phi}$\}) to move them on more likely position (\{$\bm{\phi}'$\}), according to the posterior probability. First, $M$ particles are selected randomly following the prior distribution. Then, the selected particles are moved randomly according to the multivariate distribution defined by:
\begin{equation}
p(\bm{\phi})=\sum_{i=1}^{M} w_i \frac{1}{\sqrt{(2\pi)^k |{\bm{\Sigma}}|}} \exp{(-\frac{1}{2}(\bm{\phi}-\bm{\mu}_i)^T \bm{\Sigma}^{-1} (\bm{\phi}-\bm{\mu}_i))}.
\end{equation}
The various distribution peaks are generated by displacing the original estimated pair ($\bm{\mu}$) in the direction of each $\bm{\phi}_i$, of a quantity $\bm{\mu}_i = a \bm{\mu} + (1-a) \bm{\phi}_i$. The parameter $a$ is the resampling parameter, that we set to $a=0.98$ as suggested in Ref. \cite{granade2012robust}. The covariance matrix ($\bm{\Sigma}$) is calculated by multiplying $(1-a^2)$ with the covariance matrix of the initial particles. As a result, the particles are rearranged by increasing the density around the estimated phase values. Then, the weights of new particles are set to a uniform distribution ($w_i=1/M$), and the learning process restarts. Resampling is performed when the following condition is fulfilled: $1/\sum_{i=1}^{M} w^2_i < M/2$ \cite{granade2012robust}.

\subsection{Utility function}
The utility function $U$ defines the figure of merit that is employed to tune the control parameters $\bm{\Phi}$ during the estimation process. Canonical choices for the utility function are the information gain or the quadratic loss. In our case we chose to minimize the expected variance of the posterior distribution after each step. Given the prior distribution \{$w$\}, each output $d$ of three possible cases will update the posterior following the Bayes rule, and a precise overall variance can be assigned to that specific output. The overall variance is computed by tracing the covariance matrix associated to the posterior distribution: $U(d|\{w\}) = \mathrm{Tr}[\mathrm{Cov}(\bm{\phi}|d,\{w\})]$. The expected variance $U(\bm{\Phi})$ is computed by averaging this quantity over the probability to obtain that specific output $p(d|\bm{\Phi})$. More specifically, the utility function reads: 
\begin{equation}
U(\bm{\Phi})=\sum_{i=1}^{3} p(d|\bm{\Phi}) U(d|\{w\})
\end{equation}
where $p(d|\bm{\Phi})$ is given by $p(d|\bm{\Phi}) = \sum_{i=1}^{M} w_i p(d|\bm{\phi},\bm{\Phi})$. Finally, we note that $p(d|\bm{\phi},\bm{\Phi})$ represents the likelihood of the system, which is discussed in the Sec. II. A suitable characterization of the device is thus crucial to correctly apply the algorithm. 

\subsection{Adaptive two-phase estimation experiments}
\begin{figure}[ht!]
\centering
\includegraphics[width=1.\textwidth]{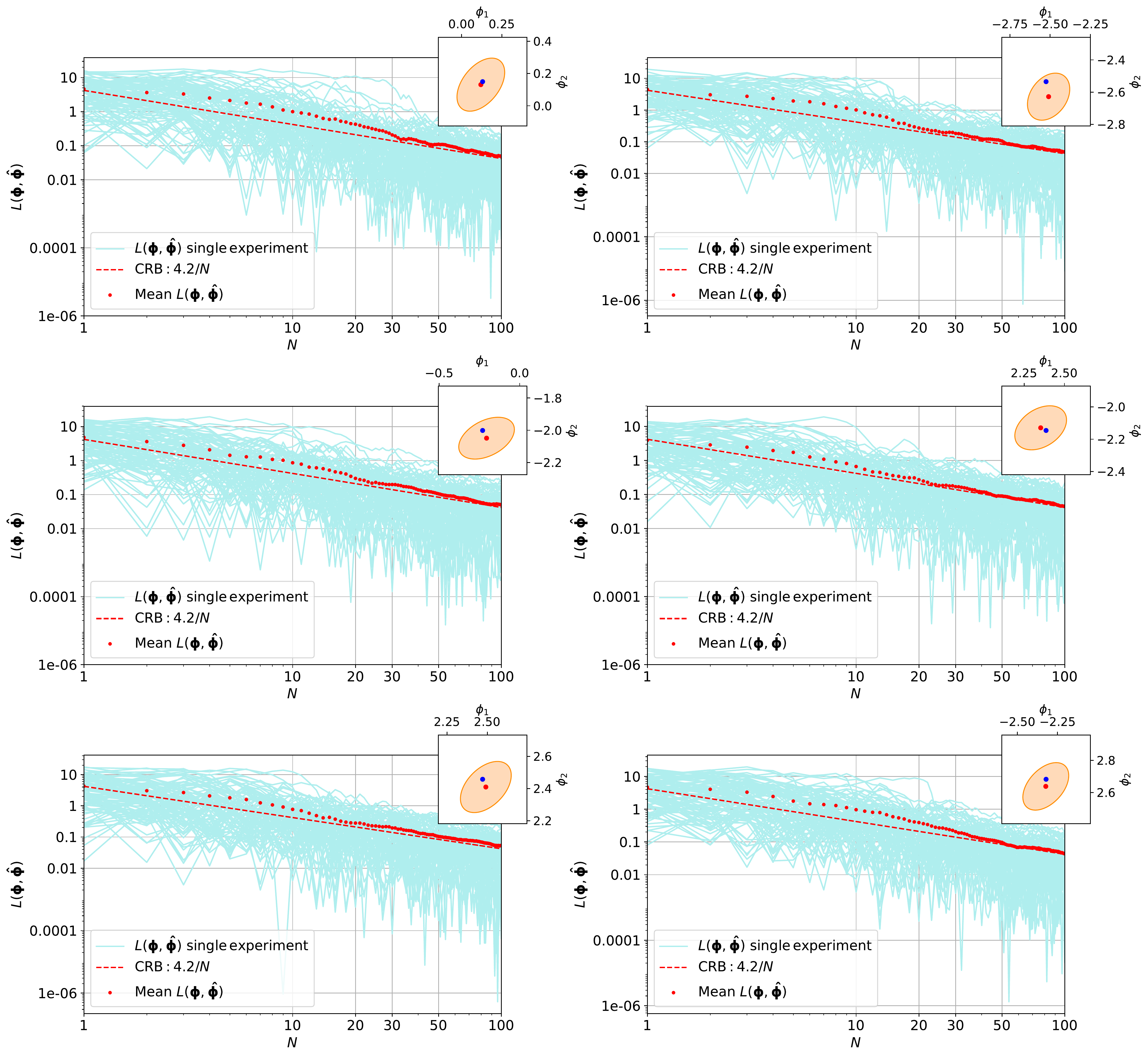}
\caption{
Examples of two-phase estimation experiments for six different pair of phases. For each pair of phases we report $L(\bm{\phi}, \bm{\hat{\phi}})$ from $N_{\mathrm{exp}}=100$ independent experiments (light blue solid lines). The average over all experiments (red dots) saturates the CRB ($\mathrm{Tr}[(\mathcal{F}_{\mathrm{exp}})^{-1}]/N=4.2/N$) (dashed lines). The inset shows the average estimated phases (red dots), the true phases to be estimated (blue dot) and the confidence interval associate to the covariance matrix (orange region).
}
\label{fig:exp}
\end{figure}

In Fig. \ref{fig:exp} we report some examples of multiphase estimation experiments performed on our device, by using the adaptive protocol. In particular, we report the overall quadratic loss $L(\bm{\phi}, \bm{\hat{\phi}})$ obtained for six different pairs, which is estimated $N_{\mathrm{exp}}=100$ independent times. The mean $L(\bm{\phi}, \bm{\hat{\phi}})$ over all $N_{\mathrm{exp}}$ runs demonstrates that the Cram\'er-Rao bound (CRB) is saturated for all pairs of phases, after a limited number of data. The CRB is computed as $\mathrm{Tr}[(\mathcal{F}_{\mathrm{exp}})^{-1}]$, where $\mathcal{F}_{\mathrm{exp}}$  is the reconstructed Fisher Information matrix of the device (see Sec. III). Our results show that all the estimated phases are close to the true set phases within one standard deviation, thus confirming the efficiency of the adopted learning algorithm. The estimation error is directly provided by the Bayesian protocol through the covariance matrix of the posterior distribution.     

\section{Likelihood estimation of the circuit}

A crucial step for the implementation of phase estimation protocols is a precise characterization of the circuit. Modeling of its action comprises 58 parameters that have to be evaluated. 

The 14 static parameters are beam-splitter transmittivities $t_i$ ($i=1,\ldots,6$), the relative phase shifts with zero applied current $\phi_{j0}$ ($j=1,2$) and $\phi_T^k$ ($k=A,B$), and the two independent visibilities with constant background noise. Conversely, the dynamic parameters are the remaining 44 response coefficients $\alpha_{ji}$ and $\alpha_{jik}^{NL}$ [Eq. (1) of the main text]. Of those parameters, 24 describe the action of the internal resistors $R_i$ ($i,j=1,\ldots,6$) and 4 describe the tritter resistors $R_j$ ($j=A,B$). Indeed, the phases $\phi_T^A$ and $\phi_T^B$ have been considered independent from the power dissipated on other resistors, since cross-talk effects with the remaining elements of the device are negligible. The last 16 parameters describe the dependence of the 8 resistors on the applied currents. We measured the three output probabilities $P(1 \rightarrow i)$ ($i=1,2,3$), injecting single photon along input 1 of the circuit. More details on the employed characterization method can be found in Ref. \cite{polino2019experimental}.

\begin{figure}[h]
\centering
\includegraphics[width=0.7\textwidth]{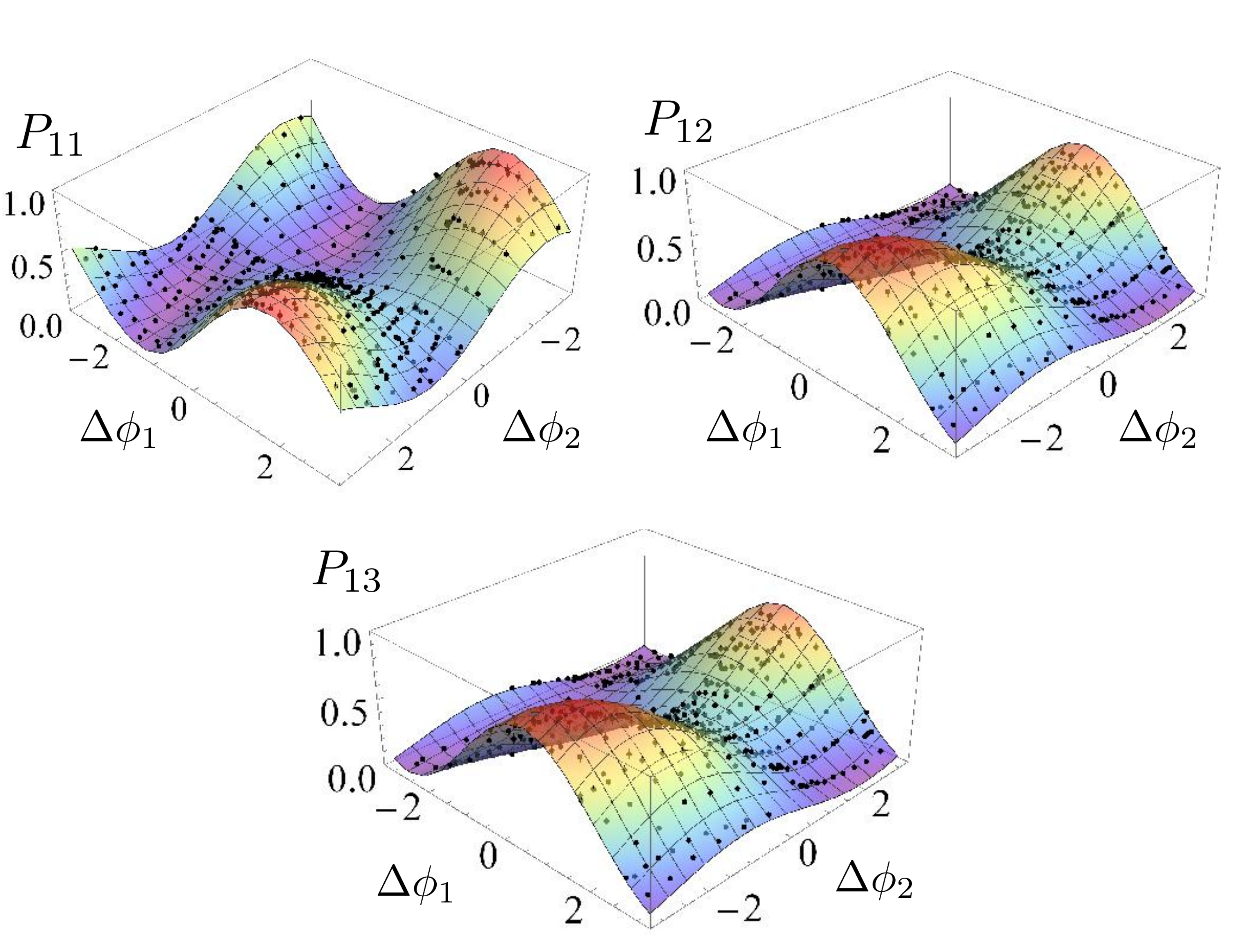}
\caption{
3D plots of output probabilities. Measured single-photon output probabilities $P(1 \rightarrow i)$ as a function of phase differences $\Delta \phi_{1}$ and $\Delta \phi_{2}$, tuned by simultaneously and independently varying the currents through resistors $R_{1}$ and $R_{2}$. Experimental data are represented by black points. Colored surfaces represent the likelihood function obtained from the characterization process. Error bars, relative to experimental points, represent standard deviations due to Poissonian statistics.}
\label{fig:plot3D}
\end{figure}

We performed different measurements of the output probabilities by fixing the currents on the resistors relative to the unknown phases ($R_3$, $R_4$ and $R_5$), and by varying those applied on the resistors associated to feedback phases ($R_1$ and $R_2$). For instance, when no current is applied on resistors $R_3$, $R_4$ and $R_5$, we estimate the chip parameters by fitting the probabilities as function of the currents applied on $R_1$ and $R_2$. The obtained value of the $\chi^2$ is 1428 with 1200 experimental points. The comparison between the fit function and the experimental data is shown in Fig. \ref{fig:plot3D}.
Note that the values of the normalized $\chi^2$ is below 1.3 for each data set. 

The probability function, that depends on the fitted parameters, represents the likelihood associated to the circuit and is the basis of the Bayesian protocol described in the main text.

\section{Fisher Information matrix}

After the characterization of all parameters, we reconstructed the Fisher Information matrix $\mathcal{F}_{\mathrm{exp}}$. Then, we optimized  the quantity $\mathrm{Tr}[(\mathcal{F}_{\mathrm{exp}})^{-1}(\Delta \phi_1, \Delta \phi_2,\phi_T^A,\phi_T^B)]$ over the phases $\Delta \phi_1, \Delta \phi_2,\phi_T^A,\phi_T^B$,  in the range of total dissipated power permitted by the circuit. Indeed, a total dissipated power greater than $1$ W could damages to the resistors. We found that the minimum value of this quantity is reached when the single photons are injected along input 1, arm 2 of the circuit is chosen as a reference, and the values of tritter and internal phases are the following: $\phi_T^A=1.49$ rad, $\phi_T^B=0.72$ rad, $\Delta \phi_1= -3.07$ rad and $\Delta \phi_2 = 0.34$ rad.

Fixing these conditions we reconstruct the Fisher Information matrix, obtaining:
\begin{equation}
    \mathcal{F}_{\mathrm{exp}}=\begin{pmatrix}0.548&-0.226\\-0.226&0.585\end{pmatrix} \qquad \qquad \mathcal{F}_{\mathrm{exp}}^{-1}=\begin{pmatrix}2.171&0.839\\0.839&2.034\end{pmatrix}.
\end{equation}
Hence, estimating the two phases with $N$ probes, the bound over the sum of the quadratic losses is: 
\begin{equation}
N L(\bm{\phi}, \bm{\hat{\phi}}) \geq\mathrm{Tr}[(\mathcal{F}_{\mathrm{exp}})^{-1}]= 4.2    
\end{equation}

\providecommand{\noopsort}[1]{}\providecommand{\singleletter}[1]{#1}%